\documentclass[a4paper,12pt]{article}
\usepackage{setspace}
\usepackage{graphicx}
\title{Anisotropic lattice deformation of InAs self-assembled quantum 
dots embedded in GaNAs strain compensating layers}
\author{N. Matsumura\thanks{Electronic mail:nmatsu@riken.jp} \\
\textit{\small Institute for Physical and Chemical Research(RIKEN), Wako 351-0198, Japan} \\
\\
S. Muto$^{a),c)}$, S. Ganapathy$^{b),c)}$ and I. Suemune$^{b),c)}$ \\
$^{a)}$\textit{\small Department of Applied Physics, Hokkaido University, Sapporo 060-8628, Japan} \\
$^{b)}$\textit{\small Research Institute for Electronic Science, Hokkaido University, Sapporo 001-0021, Japan} \\
$^{c)}$\textit{\small CREST, Japan Science and Technology Corporation, Kawaguchi 332-0012, Japan} \\
\\
K. Numata and K. Yabuta \\ 
\textit{\small Kanagawa High-Technology Foundations, Kawasaki 213-0012, Japan}}
\begin{document}
\maketitle
Lattice deformations of InAs self-assembled quantum dots, which were grown on (001) GaAs 
substrates and embedded in GaNAs strain compensating layers (SCLs), were examined with 
an ion-channeling method in Rutherford backscattering spectrometry. The channeling 
experiments demonstrated that the increase of the nitrogen concentrations in the GaNAs SCLs 
caused the indium lattice displacements along the [001] growth direction while those parallel 
to the (001) crystal plane were kept unchanged
\clearpage
	It is well-known that InAs self-assembled quantum dots (SAQDs) on (001) GaAs are 
grown via the lattice mismatch of about 7\% between InAs and GaAs. Due to their high 
crystalline quality of InAs SAQDs, a lot of studies have been performed on their applications 
to optoelectronic devices\cite{arakawa82}--\cite{leonard93}. Especially the applications of SAQDs to optical-fiber 
communications require SAQDs emission wavelengths at 1.3 $\mu$m or 1.55 $\mu$m to fit the 
minimum optical absorption bands\cite{yeh00}. Although the emission wavelength of 1.3 $\mu$m has been 
achieved by a number of groups\cite{mukai99}, the wavelength of 1.55 $\mu$m has been difficult due to the 
compressive strain induced within InAs SAQDs during the embedding processes with GaAs\cite{saito98}.
This strain issue has been coped with apparently opposite methods. One is to embed 
InAs SAQDs with InGaAs strain reducing layers (SRLs)\cite{nishi99}-\cite{tatebayashi01}, which reduce the interface 
lattice mismatch between InAs SAQDs and InGaAs SRLs by increasing indium (In) in the 
SRLs. However the higher In concentrations in the SRLs will accumulate the amount of 
overall compressive strain in the system, which may degrade the photoluminescence (PL) 
efficiencies by the possible generations of nonradiative recombination centers due to the 
excess strain. 
The other method is to embed InAs SAQDs in tensile-strained GaNAs strain 
compensating layers (SCLs)\cite{zhang02}.  This method may increase the mismatch at the InAs/GaNAs 
interfaces, but the overall average strain in the system can be minimized by compensating the 
compressive strain in the InAs SAQDs with the tensile strain in the GaNAs SCLs. Sasikala et 
al. realized 1.55 $\mu$m emission from InAs SAQDs embedded in GaNAs SCLs with the nitrogen 
(N) concentration of 2.7\%\cite{ganapathy03}. Although GaNAs usually shows degraded PL efficiencies 
with the increase of the N concentrations, the luminescence from the InAs SAQDs embedded 
in the GaNAs SCLs showed the improved efficiencies up to 5-times with the increase of the N 
concentrations\cite{ganapathy03}--\cite{zhang03}. However, the details of the strain distribution within and around the 
InAs SAQDs are not well understood.

	In this letter, the lattice deformations of InAs SAQDs embedded in GaNAs SCLs were 
examined with ion channeling in Rutherford backscattering spectrometry (RBS). The 
channeling investigations are highly sensitive to atomic displacements\cite{haga85}--\cite{feldman82}. Clear 
dependences of the anisotropic In lattice displacements on the N concentrations in the GaNAs 
SCLs will be demonstrated, and the strain relation between the InAs SAQDs and the GaNAs 
SCLs will be discussed.

	All the InAs SAQDs samples were grown on (001) GaAs substrates by metalorganic 
molecular-beam epitaxy (MOMBE). The metalorganic precursors used in this study were 
triethylgalium(TEGa), triethylindium(TEIn) , trisdimethylaminoarsenic(TDMAAs), and 
monomethylhydrazine(MMHy) for Ga, In, As, and N, respecitvely. A 300-nm-thick GaAs 
buffer layer was firstly grown at the substrate temperature of $550^{\circ}\mbox{C}$. Subsequently the 
substrate temperature was lowered to $450^{\circ}\mbox{C}$ and about 2.0 MLs of InAs were grown. A 
transition from the two-dimensional to three-dimensional growth mode, i.e., the initiation of 
the Stranski-Krastanow growth mode of the InAs layer was monitored with reflection high-
energy electron diffraction by the diffraction pattern change from streaky to spotty 
ones. A 10-nm-thick GaNAs SCL and a 10-nm-thick GaAs layer were subsequently grown at 
the same substrate temperature of $450^{\circ}\mbox{C}$. Following this sequence, three stacks of the InAs 
SAQDs layers were grown. Three samples with the N concentrations of 0.7, 1.4, and 2.65\% in 
the GaNAs SCLs were prepared. The schematic of the sample structure is shown in Fig. 1.

	A standard experimental arrangement for ion channeling was used with a tandem-type 
ion accelerator at Kanagawa High-Technology Foundations. The samples were set on a three-
axis goniometer.  2.34 MeV $\mbox{He}^{+}$ ions were used as a probe beam to investigate both [001] 
and $<110>$ channeling properties. The scattering angle and the beam spot were $160^{\circ}$ and 
1mm$\phi$, respectively. To evaluate the lattice deformation, the normalized minimum 
backscattering yield, $\chi_{min}$, which is defined as the ratio of aligned yields to random ones, was 
used. Since the relative displacement of In atoms from the GaAs host lattice increases the 
scattering of the incident ions, the increase of the minimum backscattering yield, $\chi_{min}$, 
sensitively reflects the lattice distortion due to the strain.

	One of the RBS spectra measured under the random, [001] and $<110>$ channeling 
geometries are shown in Fig. 2. The InAs SAQDs sample shown in Fig. 2 was embedded in 
the GaNAs SCL with the N concentration of 0.7\%. The inset shows In signals detected 
around the channel number of 430. The filled circles, triangles, and squares indicate the 
random spectrum, the [001] and $<110>$ spectra, respectively. Due to the channeling effect, 
drastic decreases of the backscattering yield in the [001] and $<110>$ channeling geometries 
were clearly observed compared with that in the random one. Usually the channeling in the 
$<110>$ directions is more enhanced and the backscattering yields in this direction are lower 
than those in the $<100>$ directions. However the present RBS measurements on the InAs 
SAQDs embedded in the GaNAs SCLs resulted in the reversed trend, i.e., higher 
backscattering yields in the $<110>$ channeling direction than those in the [001] direction. This 
peculiar trend was observed in all the samples measured in this study and this point will be 
discussed later.

	The N concentration dependence of the minimum backscattering yield $\chi_{min}$ for In is 
summarized in Fig. 3. This result reveals a significant difference between $\chi_{min}$ [001] and 
$\chi_{min}<110>$. Although $\chi_{min}$ [001] did not change significantly regardless of the N concentration in 
the GaNAs SCL, $\chi_{min}<110>$ showed the drastic increase with the increase of the N 
concentration in the SCL. Since the backscattering in the $<110>$ channeling direction reflects 
the In lattice tetragonal deformations, this $\chi_{min}$ [001] and $\chi_{min}<110>$ dependences demonstrate 
that the vertical distortion of the In lattices in the InAs SAQDs dominates with the increase of 
the N concentration in the GaNAs SCLs, while keeping their In lattices in the (001) plane 
nearly unchanged.

	In addition to the examination of the In lattices in the InAs SAQDs, the channeling 
properties of the Ga and As lattices around the channel number of 400 were studied, where 
the 10-nm-thick GaNAs SCLs/10-nm-thick GaAs layers burying the InAs SAQDs close to the 
sample surface mainly contribute. Figure 4 summarizes the N concentration dependence of 
the measured $\chi_{min}$. Although $\chi_{min}$ [001] for the Ga and As lattices remained almost unchanged, 
$\chi_{min}<110>$ showed the clear increase for the higher N concentrations in the GaNAs SCLs. 
This N concentration dependence is very similar to that observed for In atoms. The RBS 
measurements shown in Figs. 3 and 4 demonstrate that the deformations of both the In lattices 
in the InAs SAQDs and the Ga and As lattices in the GaNAs SCLs are mainly in the vertical 
direction to the (001) plane.

	In the present RBS measurements, the reversal of the minimum scattering yields in the 
[001] and $<110>$ channeling directions was observed compared with unstrained bulk crystal 
measurements as discussed above on Fig. 2.  This reversed trends will be attributed to the 
strain-induced lattice distortions. The tetragonal lattice distortions suggested by the results 
shown in Figs. 3 and 4 will more critically influence the $<110>$ channeling direction, which is 
inclined relative to the (001) surface, compared with the [001] direction normal to the (001) 
surface. This will make the observed reversal of the $\chi_{min}$ values relative to the unstrained 
lattices probable. 

	We have previously shown that $\chi_{min}$ measured in $<100>$ channeling directions are 
dependent on the sizes of InAs SAQDs buried near the sample surfaces, i.e., $\chi_{min}$ measured 
from samples with larger-sized dots is larger than those measured from samples with smaller-
sized dots\cite{matsumura01}. In this regard, the InAs SAQDs in the present study were prepared under the 
same conditions for all the samples.  Although there remains the possibility that the sizes and 
shapes of InAs SAQDs may change during the embedding processes, the observation of the 
nearly constant $\chi_{min}$ in the [001] direction in this study will exclude such deformation of InAs 
SAQDs during the embedding processes with the GaNAs SCLs. 

	The concept of the InGaAs SRLs\cite{nishi99}--\cite{tatebayashi01} is based on the reduced interface mismatch 
between InAs/(In)GaAs heterointerfaces. The formation of the InAs/GaNAs heterointerfaces 
in this regard will apparently increase the interface lattice mismatch.  The present observations showed 
that the In lattices in the (001) crystal planes of the InAs SAQDs were not much affected 
through the embedding processes with the GaNAs SCLs. On open InAs SAQD surfaces, however, 
surface In atoms experience the stress-free condition and the lattice extensions in both surface 
normal and lateral directions will take place. Formation of As-Ga bonds on this surface will induce 
compressive strain in the surface In-As bonds and tensile strain in the adsorbed As-Ga bonds.  
S. B. Zhang et al discussed the surface-reconstruction-enhanced solubility of N in III-V 
semiconductors based on a calculation of the substitutional energy of N atoms in binary (001) 
films\cite{zhang97}. The main mechanism to enhance the N solubility beyond the thermal equilibrium 
limit in III-V was discussed to be the reduction of the compressive strain underneath the 
surface anion dimers by the N incorporation in the sub-surface lattice sites. The present 
situation will be very similar in the sense that the N incorporation in the As sites in the 
compressively strained surface In-As bonds will reduce the compressive strain underneath 
the surface-formed As-Ga bonds. This mechanism may help to keep the overall coherent 
growth condition which keeps the (001) lattice structure unchanged.
	
	The deformations of the In lattices in the InAs SAQDs and the Ga and As lattices in 
the GaNAs SCLs in the direction normal to the (001) crystal plane and their deformation 
enhancements with the increase of the N concentrations in the GaNAs SCLs observed in this 
work will be schematically represented as shown in Fig. 5. The InAs lattice will experience 
bi-axial compression and will extend toward the direction normal to the (001) surface. 
However embedding them with GaAs layer will induce the additional compressive strain 
normal to the (001) surface. The partial replacement of the GaAs embedding layer with the 
tensile-strained GaNAs layers will shrink themselves normal to the (001) surface and this 
allows the InAs lattice to recover the expansion normal to the (001) surface as shown in Fig. 5. 
This strain release in the InAs SAQDs will explain the observed red-shift up to 1.55$\mu$m with 
the GaNAs SCLs reported in Ref.10.

	In summary, the lattice deformation of InAs SAQDs was examined with the RBS ion-
channeling method. InAs SAQDs embedded in GaNAs SCLs showed the significant increase 
of the back scattering yields in the $<110>$ channeling direction with the increase of the N 
concentrations in the GaNAs SCLs, while the backscattering yields in the [001] channeling 
direction remained nearly the same. These results demonstrated that the lattice distortions 
caused by the embedding processes of InAs SAQDs with the GaNAs SCLs are dominated in 
the direction normal to the (001) surfaces.

\clearpage
\begin{center}
Figure captions
\end{center}
\begin{enumerate}
\item[Figure 1]Schematic of InAs SAQDs sample embedded in GaNAs SCL.
\item[Figure 2]Typical RBS/channeling spectra for InAs SAQDs embedded in GaNAs SCL. Inset 
shows In signals near 430 ch.
\item[Figure 3]Normalized minimum backscattering yield ($\chi_{min}$) of In atoms as a function of the N 
concentration in the GaNAs SCL.
\item[Figure 4]Normalized minimum backscattering yield ($\chi_{min}$) of Ga and As stoms in GaNAs SCL 
as a function of the N concentration in the GaNAs SCL.
\item[Figure 5]Schematic of lattice distortion of InAs SAQDs embedded in GaAs (left) and in 
GaNAs (right).
\end{enumerate}
\clearpage
\begin{figure}
\begin{center}
\includegraphics[scale=1.0]{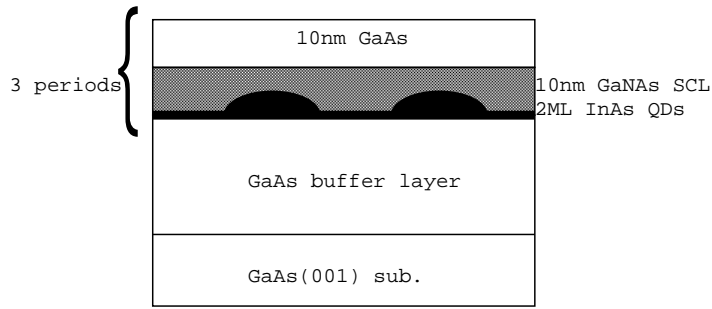}
\caption{Schematic drawing for InAs SAQDs embedded by GaNAs SCL}
\label{Fig.1}
\end{center}
\end{figure}

\begin{figure}
\begin{center}
\includegraphics[scale=1.5]{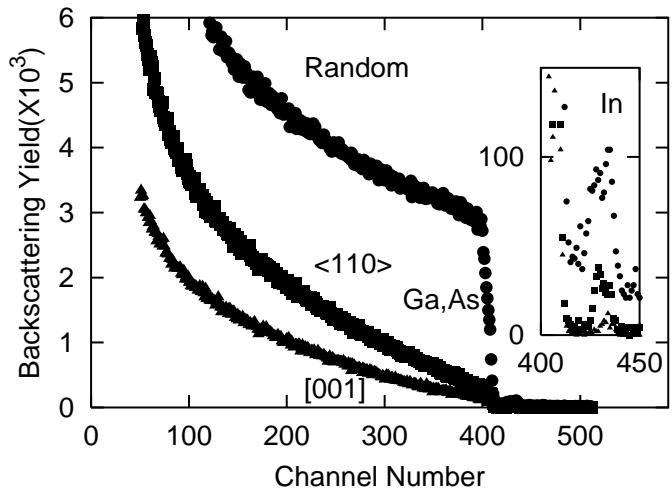}
\caption{Typical RBS/channeling spectra for InAs SAQDs embedded by GaNAs SCL. Inset shows In signals near 430ch.}
\label{Fig.2}
\end{center}
\end{figure}

\begin{figure}
\begin{center}
\includegraphics[scale=1.5]{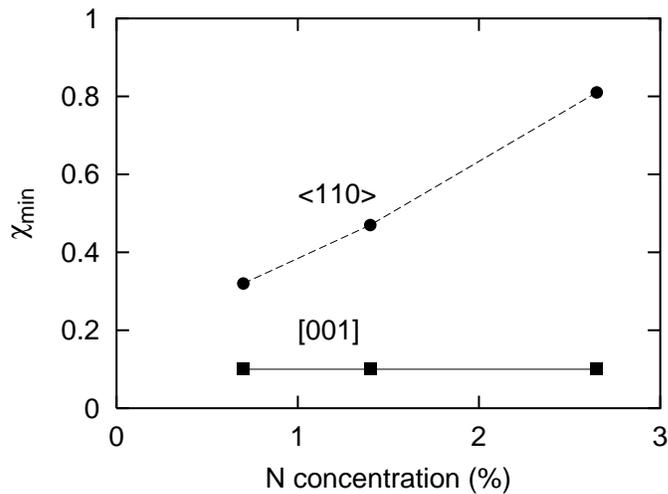}
\caption{Normalized Minimum Yield($\chi_{min}$) for In as a function of the N concentration}
\label{Fig.3}
\end{center}
\end{figure}

\begin{figure}
\begin{center}
\includegraphics[scale=1.5]{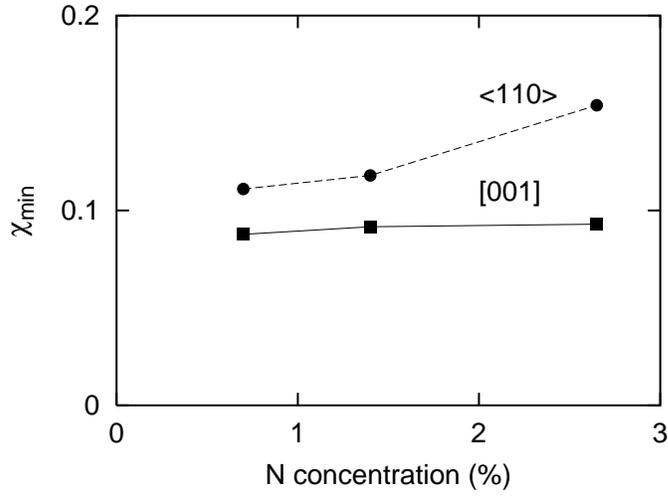}
\caption{Normalized Minimum Yield($\chi_{min}$) for around GaNAs SCL as a function of the N concentration}
\label{Fig.4}
\end{center}
\end{figure}

\begin{figure}
\begin{center}
\includegraphics[scale=0.5]{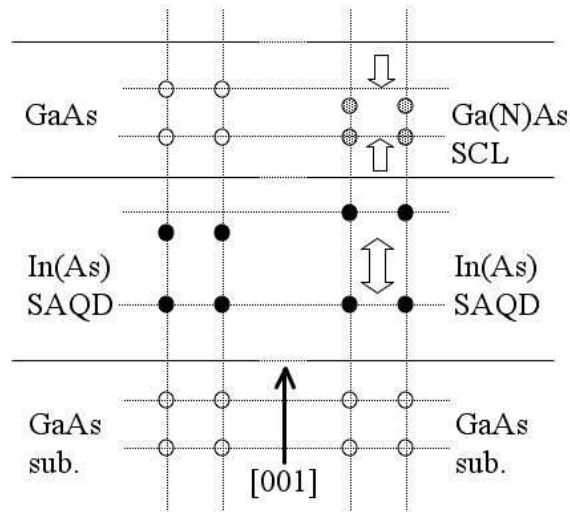}
\caption{Schematic of lattice distortion of InAs SAQDs embedded in GaAs (left) and in GaNAs (right).}
\label{Fig.5}
\end{center}
\end{figure}
\end{document}